\documentclass[twocolumn]{aastex62}

\shorttitle{Outflow Alignment in Infrared Dark Clouds}
\shortauthors{Kong et al.}

\begin{document}

\title{Widespread Molecular Outflows in the Infrared Dark Cloud G28.37+0.07: Indications of Orthogonal Outflow-Filament Alignment}

\author{Shuo Kong}
\affil{Dept. of Astronomy, Yale University, New Haven, Connecticut 06511, USA}

\author{H\'ector G. Arce}
\affil{Dept. of Astronomy, Yale University, New Haven, Connecticut 06511, USA}

\author{Mar\'ia Jos\'e Maureira}
\affil{Max-Planck-Institute for Extraterrestrial Physics (MPE), Giessenbachstr. 1, D-85748 Garching, Germany}
\affil{Dept. of Astronomy, Yale University, New Haven, Connecticut 06511, USA}

\author{Paola Caselli}
\affiliation{Max-Planck-Institute for Extraterrestrial Physics (MPE), Giessenbachstr. 1, D-85748 Garching, Germany}

\author{Jonathan C. Tan}
\affiliation{Dept. of Space, Earth and Environment, Chalmers University of Technology, Gothenburg, Sweden}
\affiliation{Dept. of Astronomy, University of Virginia, Charlottesville, Virginia 22904, USA}

\author{Francesco Fontani}
\affiliation{INAF - Osservatorio Astrofisico di Arcetri, L.go E. Fermi 5, I-50125, Florence, Italy}

\begin{abstract}
We present ALMA CO(2-1) observations toward a massive
infrared dark cloud G28.37+0.07.
The ALMA data reveal numerous molecular 
(CO) outflows with a wide range of sizes throughout the cloud. 
Sixty-two 1.3 mm continuum cores
were identified to be driving molecular outflows.
We have determined the position angle in the plane-of-sky of
120 CO outflow lobes and studied their distribution.
We find that the distribution of the plane-of-sky outflow position angles
peaks at about 100\arcdeg, corresponding to
a concentration of outflows with an approximately east-west direction.
For most outflows, we have been able to estimate the plane-of-sky 
angle between the outflow axis and the filament that harbors the
protostar that powers the outflow.  
Statistical tests strongly indicate that the 
distribution of outflow-filament orientations is consistent with
most outflow axes being mostly orthogonal to their parent filament in 3D.
Such alignment may result from filament fragmentation
or continuous mass transportation from filament to 
the embedded protostellar core. The latter is suggested
by recent numerical studies with moderately strong
magnetic fields.
\end{abstract}


\section{Introduction}

While it is generally accepted that low-mass stars
(M$\la$8M$_\odot$) form in low-mass dense molecular cores
\citep{1987ARA&A..25...23S}, the formation of massive stars
(M$\ga$8M$_\odot$) has been actively debated 
\citep{2018ARA&A..56...41M}. One of the central questions
in the study of massive star formation (SF) is how massive stars gain
their mass \citep{2014prpl.conf..149T}. One group of theories
extend the idea of low-mass SF and argue that 
massive SF begins with a massive pre-stellar core 
\citep[see reviews in][]{2007ARA&A..45..565M}.
During the past decade, surveys of massive pre-stellar cores using 
interferometry were able to find candidates with masses
up to $\sim$ 30 M$_\odot$ \citep{2013A&A...558A.125D,
2014MNRAS.439.3275W,2017ApJ...834..193K,2018ApJ...867...94K}.
However, other interferometric observations suggest that
cores may not provide enough mass for massive SF
\citep[e.g.,][]{2017ApJ...841...97S}.
Alternatively, the other group of theories emphasize 
the dominant role of clump-scale accretion
($\sim$ 10 times larger than a core)
in determining the final stellar mass. In this case,
massive SF begins with low-mass ``seeds'' and
no massive cores are needed 
\citep[see reviews in][]{2007ARA&A..45..481Z}.
In other words, in the first scenario the mass supply for massive SF
is initially contained in a small volume (core), while in the second 
scenario the mass supply is from a much larger volume.
Currently, both theories need to be tested by
further observations.

Auditing the mass budget of massive SF
is essential to understanding the origin of the stellar
initial mass function (IMF). In particular, it helps to clarify
whether the IMF originates from the so-called core mass function
\citep[CMF,][]{2010ApJ...725L..79C,2012MNRAS.423.2037H}.
Filaments are ubiquitious inside star-forming clouds
\citep{2014prpl.conf...27A}. One idea of filament-feeding 
SF was proposed based on line-of-sight observations of
velocity gradients along filaments 
\citep{2013ApJ...766..115K,2014A&A...561A..83P}.
The gradients were interpreted as mass flows
along the long axis of filaments. 
The flows were believed to funnel additional
mass supply to the SF ``hub'', providing one
mechanism that potentially solves the 
aforementioned mass-deficit problem. 
However, as pointed out by \citet{2014ApJ...790L..19F},
the interpretation of the flows in 
the Serpens South molecular cloud \citep{2013ApJ...766..115K}
is subject to projection effects.
\citet{2014ApJ...790L..19F} raised a few other 
possibilities that could also explain the 
velocity gradient along the filaments,
casting doubts on the filament-feeding idea.

To observationally
confirm that the filaments are indeed feeding SF, 
one needs to first seek evidence
that the protostellar accretion and 
the host filament are correlated. 
It is difficult to directly image the accretion in cores
due to the high resolution needed to trace the gas flow 
and the complexity of the gas kinematics at these scales.
A common assumption is that protostellar jets launched
by the rotating accretion disk of the protostar
can be used to trace the angular momentum direction
of the accretion disk 
\citep[see][and the references therein]{2014prpl.conf..451F}.
Bipolar molecular outflows driven by protostellar 
jets are much easier to trace than accretion disks
\citep{2007prpl.conf..245A,2016ARA&A..54..491B}. 
We can therefore compare the position angle between
outflows and filaments, both easily observable 
in large-scale maps of clouds, 
to search for evidence of the accretion-filament correlation.
For instance, if the outflow-filament orientation
shows a preferred angle,
we may confirm that filaments play an important
role in the protostellar accretion,
and possibly the final stellar mass.

Recently, \citet[][hereafter Stephens17]{2017ApJ...846...16S}
investigated the outflow-filament orientation
in the Perseus molecular cloud. 
Based on a statistical analysis 
of 57 protostars, Stephens17 found that
either the orientations of outflows with respect to their
natal filament is random, or are composed of a mixture of
parallel and perpendicular directions. The former may be
a result of perturbation from turbulent accretion
\citep{2015MNRAS.450.3306F,2016ApJ...827L..11O}.
Yet, a recent numerical study by \citet{2018MNRAS.473.4220L}
showed that a mostly orthogonal outflow-filament orientation
is possible in massive Infrared Dark Clouds (IRDCs).
Under the condition of moderately strong magnetic fields, 
\citet{2018MNRAS.473.4220L} showed that filaments form
perpendicular to the field
and protostellar cores inside the filaments launch
outflows that are perpendicular to the filaments.
Thus far, no strong observational evidence has been 
found to support the alignment between outflows and filaments.

IRDC G28.37+0.07\footnote{The name of ``G28.37+0.07'' is
from the MSX IRDC catalog from \citet{2006ApJ...639..227S} 
and has been used by 
\citet{2015A&A...578A..29S,2016ApJ...829L..19L,
2018ApJ...855L..25K}. However, \citet{2000ApJ...543L.157C}
originally used ``G28.34+0.06'' and this name has been used by
\citet{2006A&A...450..569P,2008ApJ...672L..33W,
2009ApJ...696..268Z,2011ApJ...735...64W,2015ApJ...804..141Z}.}
(hereafter G28) is a massive (10$^5$ M$_\odot$) IRDC at a
distance of 5 kpc. The central region of the cloud is 
dark at up to 100 $\mu$m \citep{2012A&A...547A..49R},
indicating an extremely early evolutionary stage of SF.
The dark region creates strong absorption against the
Galactic background at near and mid infrared wavelengths,
allowing researchers to derive its column density
through extinction mapping
\citep[reaching 0.7 g cm$^{-2}$,][]{2014ApJ...782L..30B}
and investigate the column density probability 
distribution function \citep{2016ApJ...829L..19L}.
At (sub-)mm wavelengths, \citet{2011ApJ...735...64W,
2015ApJ...804..141Z,2016ApJ...821L...3T,2018ApJ...867...94K} 
studied localized dense cores and outflows.
Interestingly, \citet{2011ApJ...735...64W} noticed a
hint of outflow filament alignment in the P1 clump in G28,
and later \citet{2012ApJ...745L..30W} suggested a 
perpendicular magnetic field direction in the same region.
Recently, \citet{2018MNRAS.474.3760C} found evidence
of widespread SiO emission in G28, indicating 
prevalent protostellar activities in the cloud.

In this paper, we present new ALMA CO(2-1) mosaics
that reveal widespread molecular outflows
across the IRDC G28. Our new ALMA mosaics extend
the area in G28 that was studied by the 
aforementioned papers and cover the majority of this IRDC.
The high angular resolution and high sensitivity
of the ALMA data
show strong evidence that the outflow-filament orientation
in G28 is preferentially orthogonal. The implications on
the SF process, in particular the potential connection 
between filament and protostellar accretion, 
will be discussed.

\section{ALMA Observations}\label{sec:obsalma}

The CO(2-1) data cube was obtained under the ALMA project
2015.1.00183.S (PI:Kong). It is a 86-pointing mosaic that
covers the majority of the IRDC G28.
\citet{2018ApJ...855L..25K} studied the 1.3 mm continuum 
detection under the same project. We refer the readers
to that paper for details on the ALMA observations.
The CO line data cube was cleaned using the 
{\tt tclean} task in CASA 5.1. Briggs
weighting with a robust number of 0.5 was used, resulting
in a final synthesized beam of 0.6\arcsec$\times$0.4\arcsec, 
providing a linear resolution of about 0.012 pc 
(the cloud is at a distance of 5.0 kpc). 
The maximum recoverable scale 
is $\sim$20\arcsec\ (corresponding to the 
shortest baseline of 10 k$\lambda$ with $\lambda$ being 1.3 mm), 
which is about 0.5 pc at the distance to G28.
The final cube sensitivity, with  1.3 km s$^{-1}$ wide 
velocity channels, is $\sim$ 2.5 mJy beam$^{-1}$, or 0.22 K.
As will be introduced in \S\ref{subsec:co},
we also use SiO(5-4) data to help confirm CO outflows.
The SiO data cube is from a joint de-convolution of the
data from 2015.1.00183.S and the data from
2013.1.00183.S (PI:Kong) which has the same spectral
setup (but different uv-coverage). More detailed studies
of the SiO outflows will be presented in a future paper.

\section{Analysis and Results}\label{sec:results}

\subsection{CO Outflow Definition}\label{subsec:co}

\begin{figure*}[htb!]
\centering
\epsscale{1.17}
\plotone{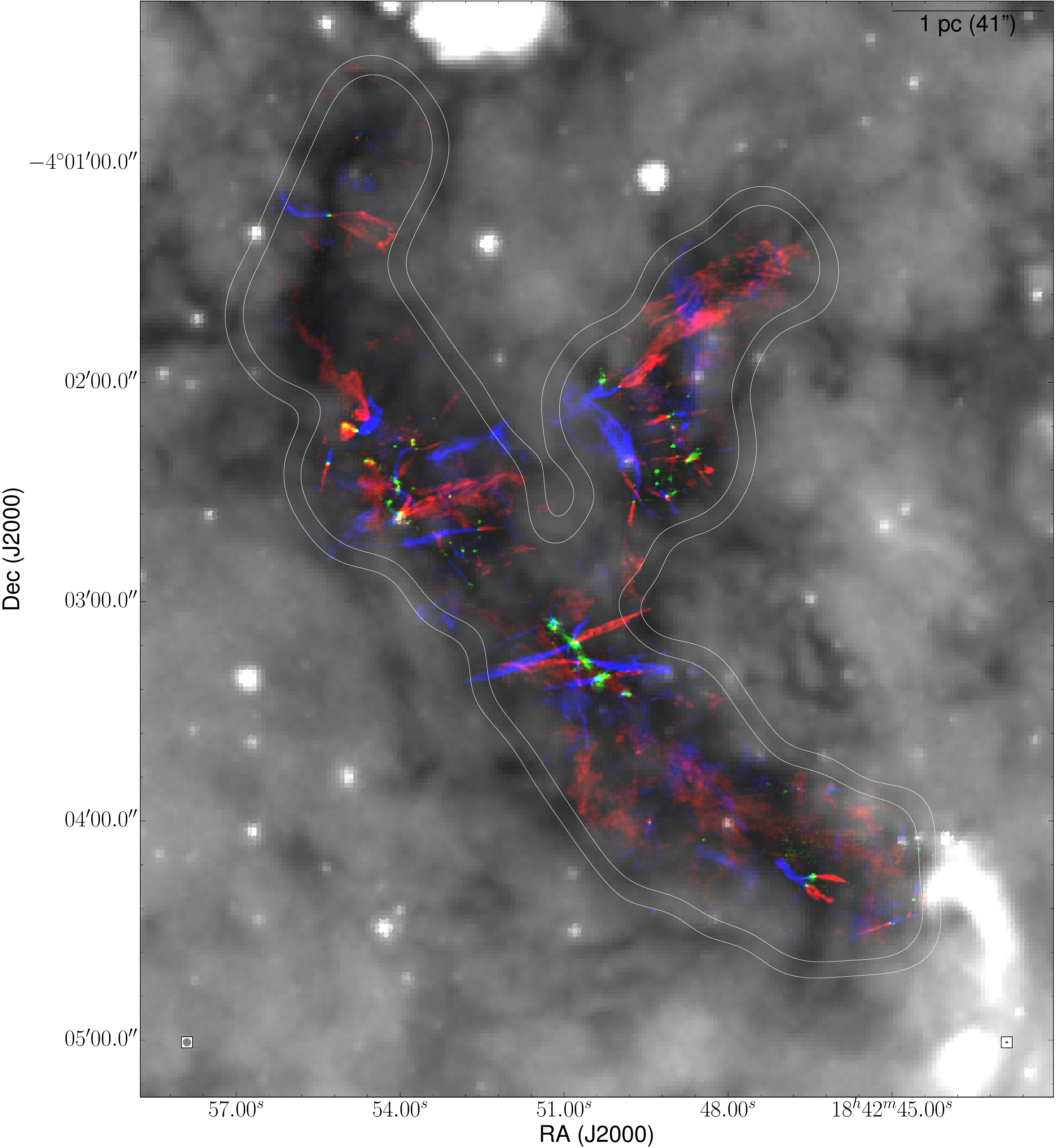}
\caption{
Map of molecular outflows in  IRDC G28. 
Background grey-scale image: Spitzer GLIMPSE 8 $\mu$m image 
\citep{2009PASP..121..213C,2003PASP..115..953B}.
The resolution beam is shown as the gray circle at the lower-left corner.
Green: ALMA 1.3~mm continuum mosaic at 0.5\arcsec\ resolution.
The green ellipse at the lower-right shows the ALMA synthesized beam.
Red: red-shifted CO(2-1) emission, integrated from 83.2 km s$^{-1}$ to 146.9 km s$^{-1}$
Blue: blue-shifted CO(2-1) emission, integrated from 11.7 km s$^{-1}$ to 75.4 km s$^{-1}$.
CO(2-1) maps have a similar spatial resolution to the continuum.
The white contours show the ALMA primary-beam responses at 30\% (outer)
and 50\% (inner). The top-right scale bar shows a scale of 1 pc.
\label{fig:outco}}
\end{figure*}

Figure \ref{fig:outco} shows a false-color image of the CO
outflows and continuum cores. 
The green color shows the 1.3 mm continuum emission.
The blue color shows the blue-shifted CO(2-1) 
integrated intensity and the red color
shows the red-shifted CO(2-1) integrated intensity.
The velocity ranges of integration
are symmetric about the cloud 
centroid velocity of 79.3 km s$^{-1}$, 
which previous studies obtained by analyzing
the spectral line data of dense gas tracers
\citep{2006A&A...450..569P,2013ApJ...779...96T,2016ApJ...821...94K}.

We detect bipolar molecular outflows that 
are distributed throughout the IRDC.
Several outflows are pc scale (possibly even
larger as some of the largest outflows 
reach the limits of the ALMA mosaic),
indicating that outflow feedback may inject
turbulent energy at $\ga$ pc scale.
Some outflows show evidence suggestive of precession.
Some show possible evidence of interaction 
with other nearby outflows.

We visually identify bipolar red-blue pairs that are
symmetric around mm continuum sources.
To produce the best results,
we made ten pairs of velocity-range integrated
intensity maps of the red-shifted and blue-shifted emission
integrated over 6.5 km s$^{-1}$ wide (5 channels) chunks,
that are symmetric around the cloud centroid velocity of
79.3 km s$^{-1}$. The ten blue-shifted
velocity-range integrated maps range from 11.7 to 75.4 km s$^{-1}$, 
while the red-shifted ones range from 83.2 to 146.9 km s$^{-1}$. 
The RMS noise for each 6.5 km s$^{-1}$ wide integrated 
intensity map is 7.3 mJy beam$^{-1}$ km s$^{-1}$,
or equivalently 0.64 K km s$^{-1}$.

With each velocity range pair, we made a false-color
image similar to Figure \ref{fig:outco}, 
and used it to visually search for bipolar CO emission. 
Then, we assigned the continuum source at the center as the 
driving source. We started with the highest velocity pair, 
which includes blue-shifted emission from 11.7 to 16.9 km s$^{-1}$ 
and red-shifted emission from 141.7 to 146.9 km s$^{-1}$, because
they only include high-velocity outflows and show 
little ``contaminating'' cloud emission.
We subsequently inspect maps of lower velocity blue-red pairs and 
add any new outflows to our list.  
We then verify our outflow catalog by carefully inspecting the 
CO emission from each channel in the cube, particularly paying
attention to  high velocity emission close to continuum sources 
to search for small outflows that might have not been clearly 
depicted in the velocity-range integrated intensity maps. 
In a few cases, we use the SiO(5-4) line data
to help determine the axes of some of the CO outflows 
that have a more complex emission structure. 
We only include unipolar outflows in our list if we 
detect high-velocity blue- or red-shifted emission 
with a morphology that suggests that it is associated with
a continuum source (e.g., if the continuum source is located
at the narrow end of an elongated high-velocity structure).

The continuum sources in this region were identified and cataloged by  
\citet[][hereafter K19]{2019ApJ...873...31K}. 
In total, 280 cores were defined in K19 using their {\it astrograph} method.
The numerical names of the cores follow
the rank of 1.3 mm continuum flux. 
K19 only studied continuum cores in the map that are within a
primary beam response $>$ 0.5 to ensure robust flux measurement.
Here, we include continuum sources that are within a
primary beam response $>$ 0.2. A small number of 
continuum sources were not defined as cores in 
K19 because of their weak emission (below 4$\sigma$)
or because their small size (smaller than half beam size). 
However, in this paper we include weak or small continuum sources
ignored by K19 which are located  at  the center of a red
and blue lobe pair or at the edge of a blue or red lobe, and thus
appear to trace the dust surrounding a protostellar outflow source.
A total of six such sources, for which we include a 'u' in their name, 
are identified. The names and coordinates of the continuum sources
associated with outflows are listed in Table \ref{tab:outflows},
columns 1-3.

Figure \ref{fig:outfil} shows the locations of the blue- and
red-shifted lobes in our map with blue and red arrows, respectively.
The arrows only indicate the position angle of the outflow axes, 
and do not represent outflow sizes (all arrows have the same length). 
We note that our outflow catalog for G28 is likely incomplete. 
Although we have gone through the CO and SiO cubes
to carefully search for outflows around continuum sources, 
we might still miss a few 
due to the complexity of the CO emission
(especially in crowded regions). 
In particular we would have likely missed  
small outflows that are within 5 km s$^{-1}$ 
of the system velocity of 79.3 km s$^{-1}$
(i.e., intrinsically low-velocity outflows or
outflows that are approximately parallel to the plane-of-sky,
yet buried in the complex cloud emission).

\subsection{Filament Definition}\label{subsec:fil}

\begin{figure*}[htb!]
\centering
\epsscale{1.15}
\plotone{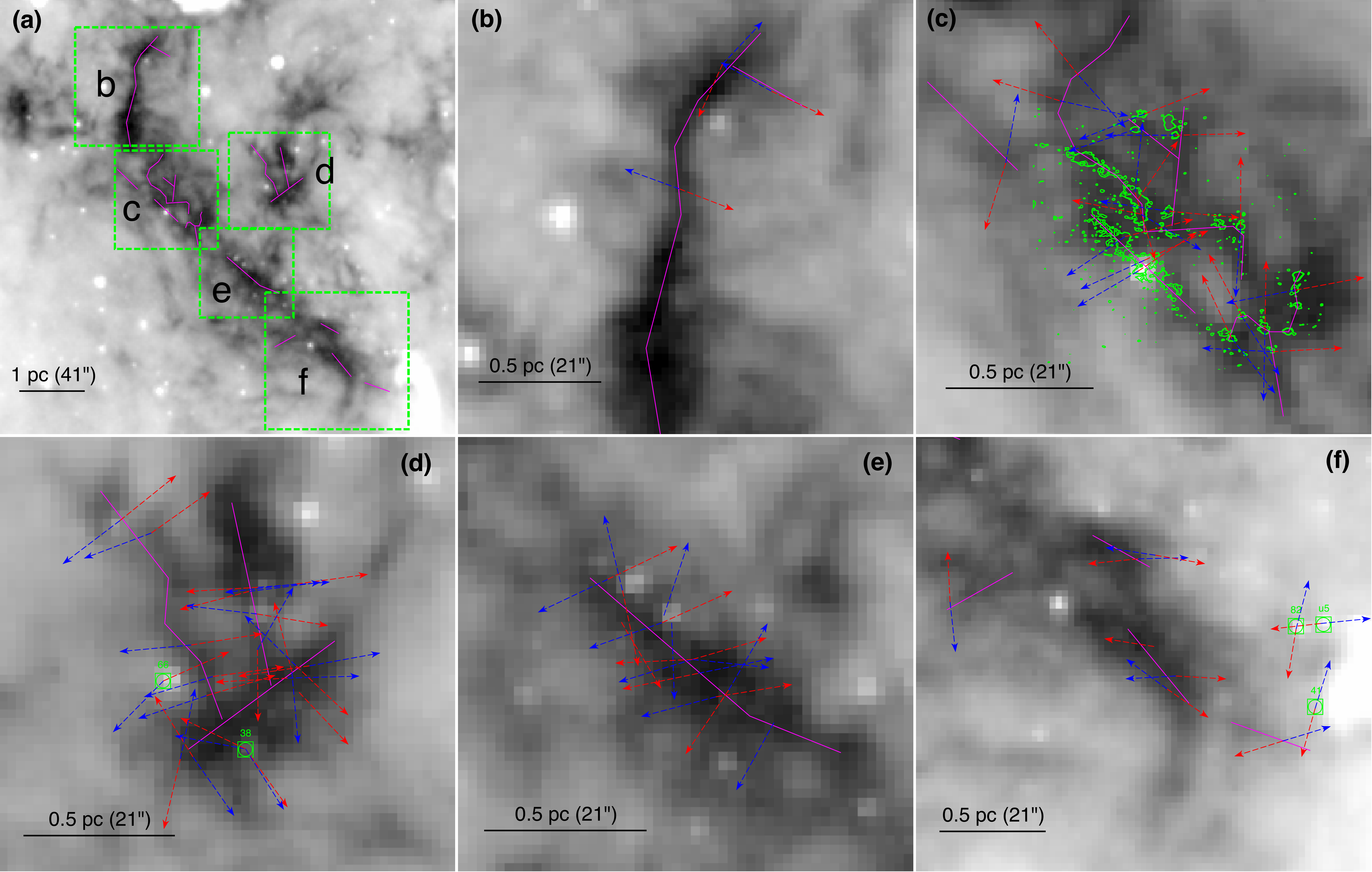}
\caption{
Summary of filaments and outflows.
Panel (a) shows an overall view. 
The magenta segments show the visually defined filaments
that pass through dark regions \citep[high column density
due to absorption, see][]{2018ApJ...855L..25K}. In panel (a),
the b-f letters and the respective dashed rectangles
represent the five sub-regions shown
in panels (b)-(f), corresponding to the five sub-regions
(``Neck'', ``Chest'', ``Claw'', ``Belly'', ``Tail'')
defined in \citet{2019ApJ...873...31K}. In panels (b)-(f),
the blue- and red-shifted outflow lobes are shown as  
blue and red dashed arrows, respectively. 
All arrows have the same length and do not represent
the lobe lengths; only the position angle on the sky is meaningful. 
The green contours in panel (c)
show the ALMA 1.3 mm continuum emission 
(contours represent 3, 4, 5$\sigma$, 
with $\sigma$ being 0.08 mJy per 0.5\arcsec\ beam).
They are used to help define the filaments where 
infrared sources break the dark filament path. 
The box-circle symbol show the protostellar cores
that cannot be associated with nearby filaments.
They are not included in the outflow-filament analysis.
A scale bar is included in each panel at the lower-left corner.
\label{fig:outfil}}
\end{figure*}

To investigate the angle between the outflow axis and the parent filament,
we define filamentary structures in G28.
Note that we only focus on filaments that harbor outflow sources 
since our goal is to study the outflow-filament angle and not to conduct 
a comprehensive study of  filaments in the IRDC.
For this purpose, we use the 8 $\mu$m image from the 
Spitzer GLIMPSE survey
\citep{2009PASP..121..213C,2003PASP..115..953B}. 
At this wavelength,
the Spitzer image gives the best spatial resolution (2\arcsec,
corresponding to $\sim$ 0.05 pc at a distance of 5 kpc) for
tracing the high surface density filaments. 
We visually follow dark paths (i.e.,  filaments in absorption)
in the cloud and overlay segments on top of the paths to 
define the filament spines (see Figure \ref{fig:outfil}).
Comparing with other techniques we find the 
visual definition to give the most reliable results.
In some of the regions we also use dust continuum emission from our 
ALMA observations to help with the filament 
definition (see Figure~\ref{fig:outfil}c). 
The 1.3 mm ALMA continuum 
is particularly helpful for defining the filament axis 
on the plane of the sky in regions where infrared sources 
in the cloud break the dark path in the Spitzer mid-IR image.
Note that the process of defining the filaments is independent
from the process of identification of outflows. 

Figure \ref{fig:outfil} shows how we defined the 
filament spines in G28. We zoom into five
sub-regions (panels b-f corresponds to the sub-regions "Neck",
"Chest", "Claw", "Belly", and "Tail" in K19, respectively), 
each containing a group of protostellar continuum cores.
In Figure \ref{fig:outfil}, the arrows represent the position 
and orientation of  outflow lobes.
In most cases the lines that define the filament 
cross the  position of the protostellar
cores that harbor the outflow sources. However, in a few cases,
cores with outflows are at the border of filaments.
In those cases, we associate
the cores to the closest filament and calculate
their relative PA. Five protostellar cores 
cannot be assigned to a nearby filament as they are relatively 
far away from a dark lane in the 8 \micron\ image
(indicated by square-circles in Figures \ref{fig:outfil} d and f).
These cores and their outflows are not included in the
outflow-filament study. 

\subsection{Outflow PA Statistics}\label{subsec:stat}

\begin{figure*}[htb!]
\centering
\epsscale{1.0}
\plotone{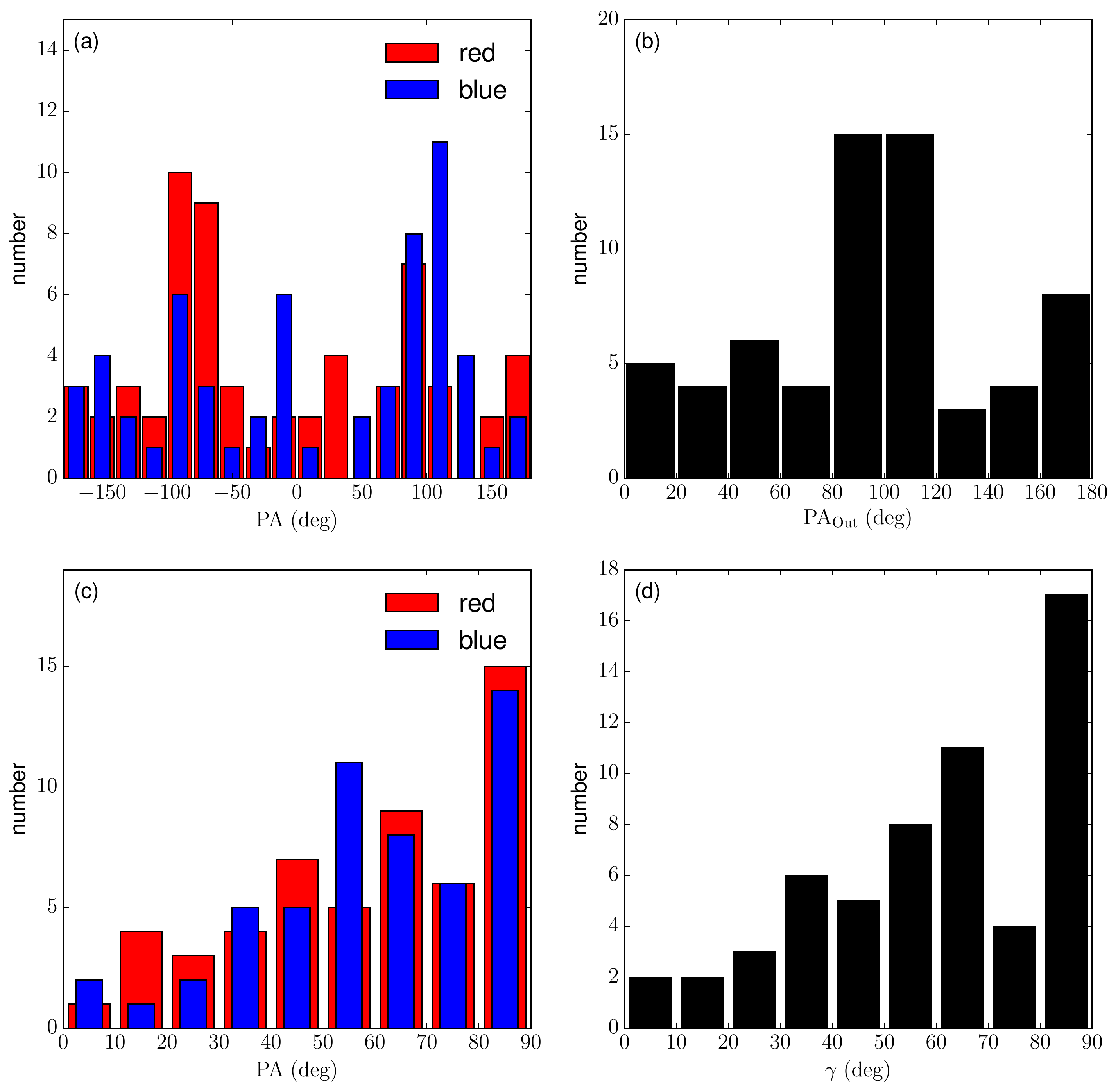}
\caption{
Outflow position angle distributions.
{\bf (a):} Position angle for blue- and red-shifted outflow lobes
separately. The angle is between -180 and 180 degrees.
{\bf (b):} Average position angle between blue- and red-shifted
lobes, i.e., PA$_{\rm Out}$. See \S\ref{subsec:stat}.
{\bf (c):} Histograms for the observed angle between 
(blue- and red-shifted) lobes and filament orientation (PA$_{\rm Fil})$.
The angle is between 0 and 90 degrees.
{\bf (d):} Histogram of $\gamma$ (the angle between PA$_{\rm Out}$ and PA$_{\rm Fil}$).
See \S\ref{subsec:stat}.
\label{fig:pa}}
\end{figure*}

\begin{figure}[htb!]
\centering
\epsscale{1.1}
\plotone{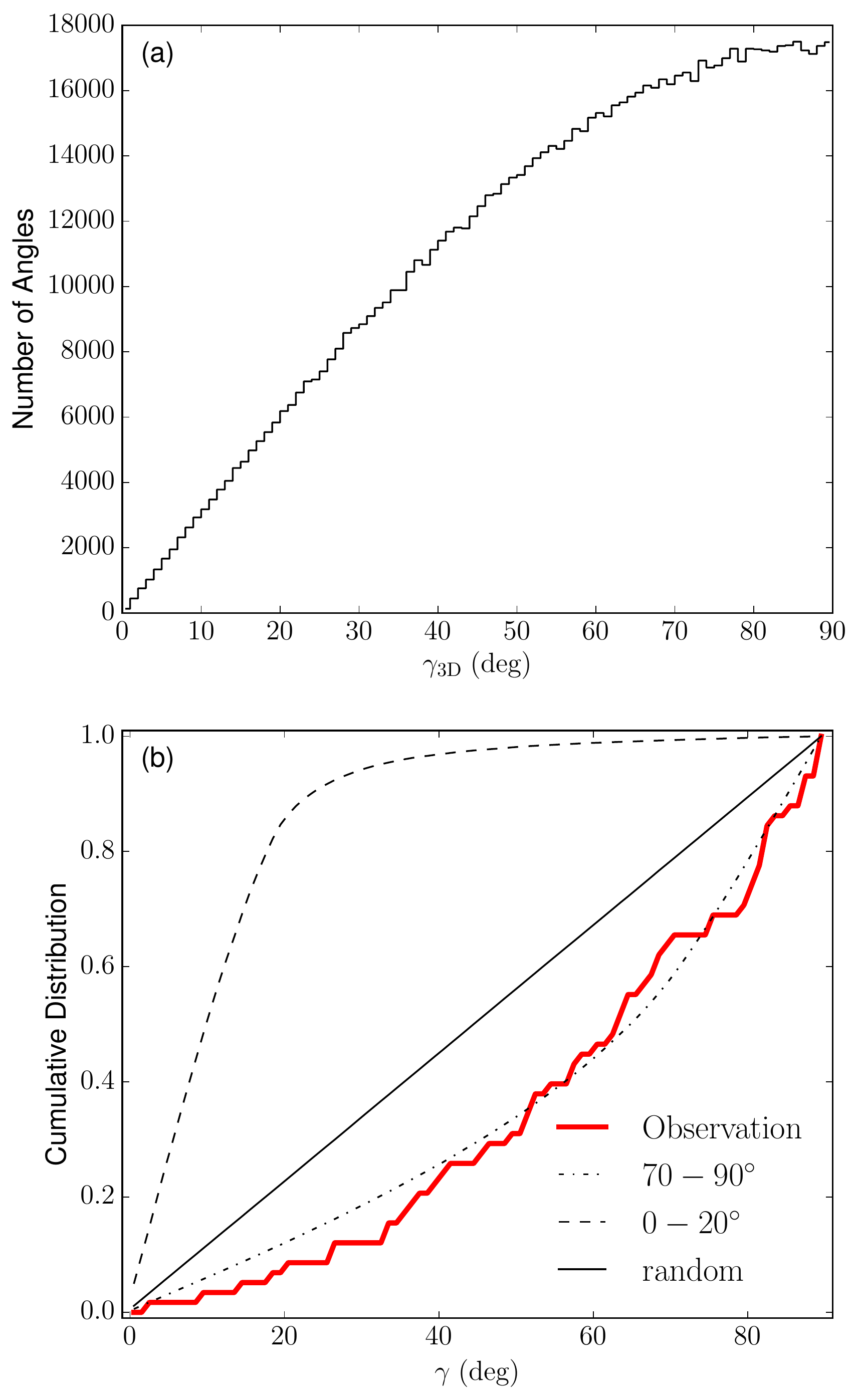}
\caption{
{\bf (a):} CDF of the simulated $\gamma_{\rm 3D}$.
See \S\ref{subsec:stat}.
{\bf (b):} CDF of the two-dimensional $\gamma$ based from the randomly-generated $\gamma_{\rm 3D}$. Three scenarios are considered
based on the distribution of $\gamma_{\rm 3D}$, i.e.,
``only parallel" (black dashed curve); ``only perpendicular"
(black dash-dotted curve); ``completely random" (black solid
curve). The CDF of observed $\gamma$ is plotted as a red curve.
See \S\ref{subsec:stat} for more details.
\label{fig:gamma}}
\end{figure}

In total, 62 continuum sources are associated with 
at least one CO outflow lobe.
Two continuum sources have double bipolar outflows.
Four continuum sources have unipolar blue-shifted outflows 
and four unipolar red-shifted outflows. 
The rest of the 62 continuum sources have bipolar outflows.
Hereafter, we follow the angle naming conventions in
Stephens17: the position angle (PA)
of the blue-shifted lobe is referred to as ``Blue PA''; 
that of the red-shifted lobe is ``Red PA''; 
the average outflow PA (see below) is denoted as ``PA$_{\rm Out}$''; 
the filament's PA  is ``PA$_{\rm Fil}$''; 
the observed angle (projected on the plane of the sky)
between PA$_{\rm Out}$ and the parent filament is ``$\gamma$''; and 
the actual (three-dimensional) difference between the outflow and filament
orientation (which is not directly observable) is  denoted as $\gamma_{\rm 3D}$
(see \S\ref{subsubsec:gamma} for more details). 
Blue PA and Red PA are defined as the angle 
from north to east (counterclockwise on the sky), ranging from -180 to 180
degree. PA$_{\rm Out}$ and PA$_{\rm Fil}$ are also defined
as the angle from north to east,
but ranging from 0 to 180 degree. 
$\gamma$ and $\gamma_{\rm 3D}$ have values ranging from 0 to 90 degree.
Blue PA, Red PA, PA$_{\rm Out}$, PA$_{\rm Fil}$, and
$\gamma$ for all the CO outflow lobes are shown in 
Table \ref{tab:outflows} columns 4 to 8, respectively.

\subsubsection{Distribution of PA$_{\rm Out}$}\label{subsubsec:paout}

First, we investigate the distribution of PA$_{\rm Out}$.
A striking feature in Figure \ref{fig:outco}
is  that a large number of the  outflows seem to 
have an approximate  east-west orientation.
To confirm this, in Figure \ref{fig:pa} panel (a), we plot 
the distribution of Blue PA and Red PA
(shown in corresponding colors). It is clear that
the histograms peak at  $\sim$ 100\arcdeg\ 
and  $\sim$ -80\arcdeg, for Blue PA and  Red PA, respectively, 
corresponding to the apparent east-west direction for most lobes mentioned above. 
We follow Stephens17 and derive
the PA$_{\rm Out}$ by adding 180\arcdeg\ to the outflow lobes with negative PA
and averaging the Blue PA and Red PA for outflows with two lobes. For   
outflows with one lobe, we just use the orientation 
of that lobe as the PA$_{\rm Out}$ (adding 
180\arcdeg\ if negative). Figure \ref{fig:pa}b
shows the distribution of PA$_{\rm Out}$ values. The histogram
shows a peak at PA$_{\rm Out}$ $\sim$ 100\arcdeg,
which is consistent with the visual examination of 
Figure \ref{fig:outco}, mentioned above.
If the three-dimensional orientation of outflows is random, then we would expect a
uniform distribution of PA$_{\rm Out}$ in two dimensions. 
Following Stephens17, we carry out Anderson-Darling (AD) tests
between the distribution of PA$_{\rm Out}$ and a uniform distribution.
The p-value of the AD test is 0.15, indicating PA$_{\rm Out}$
is unlikely drawn from a uniform distribution.
This suggests that the 3D outflow position angle is not
randomly distributed.

\subsubsection{Outflow-filament Orientation}\label{subsubsec:gamma}

Second, we investigate the relative orientation of the
outflows with respect to their parent filament ($\gamma_{\rm 3D}$).
Figure \ref{fig:pa}c shows the histograms of the observed angle 
between outflow lobes and their parent filament on the plane of the sky. 
Interestingly, the histograms show a higher number for
PA $>$ 70\arcdeg\ for both blue- and red-shifted
outflow lobes. In fact, this is readily seen in Figure \ref{fig:outfil}
where many arrows show large PA relative to the filaments.
In Figure \ref{fig:pa}d, we present the histogram of $\gamma$ (i.e., the angle between
PA$_{\rm Out}$ and PA$_{\rm Fil}$ on the plane of the sky). Again,
following Stephens17, we calculate $\gamma$ as
\begin{equation}\label{eq:gamma}
\gamma = \rm MIN\{|PA_{\rm Out}-PA_{\rm Fil}|, 180\arcdeg-|PA_{\rm Out}-PA_{\rm Fil}|\}, 
\end{equation}
where $\gamma$ is between 0 and 90\arcdeg\ (see equation 2 of Stephens17). 
In Figure \ref{fig:pa}d, it is clear that    
there is a significantly higher number of outflows 
with angles with respect to their parent filament $\gamma$
with values between 60\arcdeg\ and 90\arcdeg\ (32), 
compared to outflows with $\gamma$ between 30\arcdeg\ and 60\arcdeg\ (19),
and between 0\arcdeg\ and 30\arcdeg\ (7).

We emphasize that a large concentration of values of $\gamma$ 
close to 90\arcdeg\ does not necessarily mean 
that outflows tend to be perpendicular to the filaments in three-dimensions.
The measured values of $\gamma$ (from our images)
represent a projection of the actual three-dimensional angle
between outflow axes and their parent filament ($\gamma_{\rm 3D}$) on the plane of the sky.
In other words, a value of $\gamma = 90$\arcdeg\
can be the projection on the plane of the sky of any $\gamma_{\rm 3D}$
(see detailed discussions by Stephens17). 

Here we follow Stephens17 and carry out a Monte Carlo simulation 
that projects randomly generated $\gamma_{\rm 3D}$ onto the plane of the sky 
to produce a distribution of values of $\gamma$, 
and compare them with our observations.
The detailed methodology was described in Appendix A of Stephens17.
Here we briefly recap the process. We randomly generate $N$ pairs of unit
vectors with their 3D PA following the uniform distribution. Then we calculate  
$\gamma_{\rm 3D}$, as the angle between the two unit vectors. Next, we project
the two vectors onto the y-z plane and calculate their angle
(equivalent to $\gamma$). 
Finally, we calculate the cumulative distribution function (CDF) of $\gamma$
and compare with observations. We use $N$=10$^6$ (the same as
Stephens17) and reproduce their $\gamma_{\rm 3D}$ CDF,
shown in Figure \ref{fig:gamma}(a) (compared with Figure 14 in Stephens17).

We also follow Stephens17 and consider three scenarios:
(1) $\gamma_{\rm 3D}$ with values only from 0\arcdeg\ to 20\arcdeg; 
(2) $\gamma_{\rm 3D}$ with values only from  70\arcdeg\ to 90\arcdeg;
(3) $\gamma_{\rm 3D}$ with values from 0\arcdeg\ to 90\arcdeg. 
The first scenario corresponds to the ``only parallel"
case in Stephens17; the second the ``only perpendicular" case;
the third ``completely random". We plot the corresponding 
$\gamma$ CDF for each of these three scenarios in Figure \ref{fig:gamma}(b)
(analogous to Figure 8 in Stephens17).
It is immediately obvious that $\gamma$ does not follow the
``only parallel'' scenario.  
Quite interestingly, the observed $\gamma$ CDF appears to be
consistent with the ``only perpendicular" scenario. That is, in three-dimensional space
outflows preferentially have axes that approximately perpendicular to  the orientation of
their parent filaments, contrary to what was found in Perseus by Stephens17. 
A two-sample AD test between the observed $\gamma$ CDF and
simulated $\gamma$ CDF for the ``completely random'' scenario
(middle plot in Figure \ref{fig:gamma}b) gives a p-value of 6.5$\times$10$^{-5}$.
It is therefore very unlikely that the underlying (three-dimensional) distribution
of angles between outflows and filaments is completely random.
On the other hand, a similar two-sample AD test comparing  the observed $\gamma$ with the
simulated $\gamma$ CDF for the ``only perpendicular'' scenario gives a p-value of 0.53, 
meaning that the hypothesis  that the  underlying angles
between outflows and filaments are mostly perpendicular cannot be rejected. 
Given our results we conclude that it is more likely that in G28
outflows are perpendicular to their parent filament rather than being
parallel to, or having a random orientation with respect to, their parent filament.

\startlongtable
\begin{deluxetable*}{rccccccc}
\tabletypesize{\scriptsize}
\tablewidth{0pt} 
\tablecaption{Outflows Properties \label{tab:outflows}}
\tablehead{
\colhead{Source} & \colhead{RA}& \colhead{DEC} & \colhead{Blue PA} & \colhead{Red PA} & \colhead{PA$_{\rm Out}$} & \colhead{PA$_{\rm Fil}$} & \colhead{$\gamma$} 
} 
\colnumbers
\startdata 
      1 &  280.69353 &   -4.07084 &    93 &   -93 &    90 &    40 &    50 \\
  2\_p1 &  280.71076 &   -4.05449 &   -96 &    93 &    88 &    49 &    39 \\
  2\_p2 &  280.71076 &   -4.05449 &   107 &   -74 &   107 &    49 &    58 \\
      3 &  280.71328 &   -4.05201 &    14 &  -168 &    13 &    49 &    36 \\
      4 &  280.72522 &   -4.04092 &     - &   -34 &   146 &    41 &    75 \\
      5 &  280.71186 &   -4.05317 &   109 &   -66 &   112 &    49 &    63 \\
      7 &  280.70419 &   -4.03817 &   -30 &  -180 &   165 &    12 &    27 \\
      8 &  280.70950 &   -4.03319 &   127 &   -51 &   128 &    38 &    90 \\
      9 &  280.72832 &   -4.03704 &  -102 &    73 &    75 &   158 &    83 \\
     10 &  280.72574 &   -4.04240 &   124 &   -76 &   114 &    46 &    68 \\
     11 &  280.71151 &   -4.05317 &  -178 &     - &     2 &    49 &    47 \\
     12 &  280.70946 &   -4.05559 &   -30 &   146 &   148 &    49 &    81 \\
     13 &  280.72497 &   -4.04333 &   114 &   -65 &   115 &    46 &    69 \\
     15 &  280.70303 &   -4.03926 &   -88 &    93 &    92 &   126 &    34 \\
     19 &  280.73045 &   -4.03949 &   -10 &   163 &   166 &    45 &    59 \\
     21 &  280.70451 &   -4.03592 &   -83 &   105 &   101 &    12 &    89 \\
     22 &  280.73035 &   -4.02063 &    70 &  -111 &    69 &     5 &    64 \\
     23 &  280.70286 &   -4.03902 &  -175 &    14 &    10 &   126 &    64 \\
     33 &  280.69402 &   -4.07155 &    54 &  -125 &    55 &    40 &    15 \\
     35 &  280.70992 &   -4.05583 &   105 &   -81 &   102 &    49 &    53 \\
     37 &  280.72021 &   -4.04612 &   179 &    -1 &   179 &   133 &    46 \\
 38\_p1 &  280.70465 &   -4.04199 &    80 &  -144 &    58 &     - &     - \\
 38\_p2 &  280.70465 &   -4.04199 &  -148 &    65 &    48 &     - &     - \\
     40 &  280.72511 &   -4.04151 &  -121 &    79 &    69 &     5 &    64 \\
     41 &  280.68586 &   -4.07262 &   -17 &   164 &   163 &     - &     - \\
     44 &  280.72117 &   -4.04202 &   176 &     0 &   178 &    49 &    51 \\
     45 &  280.70717 &   -4.04238 &   -12 &   167 &   167 &   126 &    41 \\
     46 &  280.72534 &   -4.03817 &   108 &     - &   108 &    51 &    57 \\
     48 &  280.70421 &   -4.03679 &    83 &   -99 &    82 &    12 &    70 \\
     49 &  280.72399 &   -4.04140 &    82 &   -95 &    83 &   174 &    89 \\
     54 &  280.69561 &   -4.06440 &   -89 &    96 &    93 &    60 &    33 \\
     56 &  280.70218 &   -4.03879 &   -80 &    98 &    99 &   126 &    27 \\
     58 &  280.72112 &   -4.04554 &  -148 &    28 &    30 &   101 &    71 \\
     60 &  280.70417 &   -4.03576 &   -86 &    93 &    94 &    12 &    82 \\
     61 &  280.71179 &   -4.05291 &   -18 &     - &   162 &    49 &    67 \\
     62 &  280.71901 &   -4.04451 &   101 &   -79 &   101 &    10 &    89 \\
     63 &  280.72522 &   -4.04376 &   120 &   -56 &   122 &    46 &    76 \\
     65 &  280.70829 &   -4.03367 &   111 &   -55 &   118 &    38 &    80 \\
     66 &  280.70782 &   -4.03937 &   135 &   -66 &   125 &     - &     - \\
     71 &  280.72375 &   -4.03840 &    90 &   -89 &    91 &   174 &    83 \\
     75 &  280.70597 &   -4.06677 &  -174 &     3 &     4 &   120 &    64 \\
     76 &  280.72152 &   -4.04641 &  -142 &    25 &    32 &   160 &    52 \\
     79 &  280.70668 &   -4.04230 &  -145 &    32 &    33 &   126 &    87 \\
     82 &  280.68691 &   -4.06826 &   -15 &   171 &   168 &     - &     - \\
     84 &  280.68766 &   -4.07442 &   -73 &   107 &   107 &    70 &    37 \\
     85 &  280.71056 &   -4.05499 &   -79 &   102 &   102 &    49 &    53 \\
     94 &  280.72829 &   -4.01469 &   -43 &   156 &   146 &   137 &     9 \\
    106 &  280.72497 &   -4.03756 &   111 &   -69 &   111 &    51 &    60 \\
    107 &  280.69466 &   -4.06420 &    81 &  -100 &    81 &    60 &    21 \\
    109 &  280.71332 &   -4.05315 &     - &  -149 &    31 &    49 &    18 \\
    134 &  280.72534 &   -4.04065 &    -6 &  -164 &     5 &    41 &    36 \\
    149 &  280.70270 &   -4.03879 &    45 &  -133 &    46 &   126 &    80 \\
    153 &  280.70616 &   -4.03994 &   109 &   -73 &   108 &    20 &    88 \\
    169 &  280.71381 &   -4.05161 &   115 &   -64 &   116 &    49 &    67 \\
    185 &  280.71992 &   -4.04694 &    87 &   -87 &    90 &    10 &    80 \\
    214 &  280.70783 &   -4.05676 &   151 &     - &   151 &    68 &    83 \\
    237 &  280.69473 &   -4.06922 &     - &    81 &    81 &    40 &    41 \\
    264 &  280.70258 &   -4.03981 &     - &  -137 &    43 &   126 &    83 \\
     u1 &  280.72596 &   -4.01598 &    60 &  -115 &    63 &    60 &     3 \\
     u2 &  280.72757 &   -4.03604 &  -138 &    39 &    40 &   128 &    88 \\
     u3 &  280.70268 &   -4.03563 &    97 &   -82 &    97 &    12 &    85 \\
     u4 &  280.70675 &   -4.03799 &    95 &   -81 &    97 &    43 &    54 \\
     u5 &  280.68556 &   -4.06795 &   -84 &    96 &    96 &     - &     - \\
     u6 &  280.70591 &   -4.03920 &   106 &   -82 &   102 &    20 &    82 \\
\enddata
\tablecomments{Columns 1-3 are introduced in \S\ref{subsec:co}. 
Columns 4-8 are defined in \S\ref{subsec:stat}. 
Starting from column 2, all columns are in unit of degree.}
\end{deluxetable*}

\section{Discussion}

The non-random outflow-filament angle in IRDC G28 suggests
that the angular momentum of the protostellar accretion disk is
correlated with the host filament. 
Then, two possible scenarios may explain this. In one scenario, the
angular momentum direction is determined during
the core fragmentation/formation and maintained for some time.
In the other scenario, the angular momentum is
a result of continuous mass accretion
onto the core and then the accretion disk. Qualitatively, in the first
scenario, the core is relatively isolated and no significant mass is
deposited onto the core from the environment. In the second scenario,
the core is constantly acquiring mass from its host filament.

Recently, \citet{2018MNRAS.473.4220L}
studied the formation of stellar clusters in a magnetized, 
filamentary IRDC, including 
early outflow feedback from the protostars. Since they were not 
able to resolve the disk formation scales, they adopted the direction 
of the net angular momentum at scales of several hundreds of AU as
the outflow launching direction. Under these conditions, their 
results showed that outflows have a preferred orientation, which 
is perpendicular to the large-scale filament, as our observations
suggest. In a follow-up paper, \citet{2019arXiv190104593L}
showed that the outflow-filament angle is determined
by the constant mass accretion from the filament to the embedded core,
consistent with the second scenario stated above.
These results highlight the importance of filament-core-disk accretion
in SF. We note that in their simulation a moderately strong 
magnetic field is used as an initial condition. 
Such dynamically important fields in IRDCs are reported by 
recent observations \citep[e.g.,][]{2015ApJ...799...74P}.
The filaments form perpendicular to the field, 
which results in the alignment of outflows with the direction of the original magnetic fields 
on the plane-of-the-sky. Following this result,
our finding of the east-west outflow alignment in G28 (\S\ref{subsubsec:paout}) 
may indicate that the magnetic field in this cloud has an east-west orientation.
Future dust polarization measurements may confirm this.

Although we have not specifically studied the outflow size distribution,
we note that their typical sizes based on Figure \ref{fig:outco} are
0.1-1 pc. For an outflow lobe of 0.5 pc in length, 
and assuming a velocity of 100 km s$^{-1}$
\citep[the velocity of the driving jet,][]{2016ARA&A..54..491B},
the timescale is about 5000 years. This implies that
the outflow-filament alignment has been 
approximately stable for roughly the past few thousand years
(except for those showing precession-like PA variations). 
It remains to be seen how long can the cloud maintain the
filament-core-disk accretion that possibly gives rise to the 
outflow-filament alignment. Such accretion process is important
because it provides mass to the protostar from outside the core,
thus having impact on the stellar initial mass function.
More specifically, if the final stellar mass is dominated
by the filament accretion, then the CMF-IMF relation would
be questionable. On the other hand, if the filament accretion
is limited, then it is possible that there might be a 
statistically significant relation between the 
CMF and the IMF.

Several observational studies have shown that the direction of
the angular momentum at core/disk scales
is not correlated with the filamentary structure 
\citep[][Stephens17]{2009A&A...496..153D,2016PASJ...68...24T}. 
Possible causes are protostellar multiplicity 
\citep{2016ApJ...820L...2L,2016ApJ...827L..11O},
and/or an underlying bimodal distribution, with
one sub-population of outflow directions that is preferentially 
parallel and another that is perpendicular to filaments (Stephens17).
It would be useful to carefully examine any differences
between the clouds to investigate whether differences in outflow-filament
alignment is caused by the cloud environment or evolutionary stage.

\section{Summary and Conclusions}

In this paper, we have investigated the 
distribution of outflow position angles in the
infrared dark cloud G28.37+0.07. 
We have defined molecular outflows
in the CO(2-1) cube observed with ALMA. 
In total, we have identified
CO outflows from sixty-two 1.3 mm continuum cores.
Fifty-four cores have bipolar outflows
and two of them show double bipolar outflows.
Four cores have blue-shifted unipolar outflows 
and four have red-shifted unipolar outflows.
We have studied the statistics of the 
(CO) outflow position angles.
In particular, we have found that the 
outflow position angles in G28.37+0.07
are not randomly oriented. Albeit spread 
over the extent of the dark cloud, the plane-of-sky
outflows preferentially align at a
position angle of $\sim$ 100\arcdeg, roughly
corresponding to an east-west direction.

We have visually defined filamentary structures along the dark
paths in the Spitzer 8 $\mu$m near infrared image. The dark paths
correspond to high mass surface density filamentary structures.
At the location of each protostellar core that drives an outflow,
we have measured the filament position angle and compared
it with the position angle of the outflow.
In G28.37+0.07 we have observed a higher number of 
continuum sources with an observed (projected on 
the plane of the sky) angle between the outflow axis 
and the parent filament close to 90\arcdeg.
We have carried out Monte Carlo simulations of the 
three-dimensional orientation between the outflow 
and filament and projected it onto two dimension
to mimic plane-of-the-sky observations. 
A comparison of the cumulative distribution
functions of the observed and simulated plane-of-sky
outflow-filament angles shows that outflows are
preferentially perpendicular to filaments. 

The finding is consistent with the recent numerical work
\citep{2018MNRAS.473.4220L}, which supports the physical picture
that protostellar accretion is fed by gas flow along filaments
in infrared dark clouds, highlighting the importance of 
filamentary structures in SF.
On the other hand, studies in, e.g., the Perseus cloud
\citep{2017ApJ...846...16S}, have not shown such alignments.
We now need to understand whether these two seemingly 
contrasting results indicate there are  two
intrinsically different modes of SF or merely
an evolutionary effect. Since G28.37+0.07 is massive,
infrared-dark, and probably forming the
first generation stars, it provides an excellent archetype
to be compared with other well-known star-forming clouds at 
different evolutionary stages.

\acknowledgments 
We thank the anonymous referee for helpful comments.
We thank Pak Shing Li for fruitful discussions.
SK and HGA were funded by NSF award AST-1140063
and AST-1714710 while conducting this study.
This paper makes use of the following ALMA data:
ADS/JAO.ALMA\#2013.1.00183.S and ADS/JAO.ALMA\#2015.1.00183.S. 
ALMA is a partnership of ESO
(representing its member states), NSF (USA) and NINS (Japan), together
with NRC (Canada), NSC and ASIAA (Taiwan), and KASI (Republic of
Korea), in cooperation with the Republic of Chile.  The Joint ALMA
Observatory is operated by ESO, AUI/NRAO and NAOJ.  The National Radio
Astronomy Observatory is a facility of the National Science Foundation
operated under cooperative agreement by Associated Universities, Inc.
This work is based [in part] on observations made with the Spitzer 
Space Telescope, which is operated by the Jet Propulsion Laboratory, 
California Institute of Technology under a contract with NASA.

\software{Astropy \citep{Astropy-Collaboration13}, Numpy \citep{numpy}, APLpy \citep{Robitaille12}, Matplotlib \citep{matplotlib}, Glue \citep{2015ASPC..495..101B,robitaille_thomas_2017_1237692}, SAOImageDS9 \citep{2003ASPC..295..489J}}

\facility{ALMA,Spitzer};

\end{document}